\documentstyle[11pt,paspconf,psfig]{article}

\begin{document}

\font\sixrm=cmr6 \font\sixi=cmmi6 \font\sixit=cmti8 at 6pt
\vbox{\hsize=.5\hsize \raggedright
       \sixit\baselineskip=7.2pt \noindent
	UC Santa Cruz Workshop on Galactic Halos \\
        Ed.\ D. Zaritksy}

\title{The Shapes and Sizes of Elliptical Galaxy Halos from X-Ray
Observations}

\author{David A. Buote}
\affil{Institute of Astronomy, Madingley Road, Cambridge CB3 0HA,
U.K.} 

\author{Claude R. Canizares}
\affil{Department of Physics and Center for Space Research 37-241,
Massachusetts Institute of Technology, 77 Massachusetts Avenue,
Cambridge, MA 02139}

\begin{abstract}
We review the theory of how the shapes of the X-ray isophotes probe
both the shape and radial distribution of gravitating matter in
elliptical galaxies in a way that is more robust than the traditional
spherical approach.  We summarize and update the previous analyses of
X-ray observations of NGC 720 and NGC 1332 and describe preliminary
results for NGC 3923.
\end{abstract}



\section{Introduction} 

The soft X-ray band is particularly useful for studying the mass
distributions of elliptical galaxies, because the emission over
energies $\sim 0.5-2$ keV arises primarily from hot gas with
temperature $T\sim 10^7$ K; i.e. for the more massive ellipticals
having ratios of X-ray to blue-band luminosity $\log_{10}L_x/L_B>\sim$
$-2.7$ (Canizares, Fabbiano, \& Trinchieri 1987). The small mean free
path ($\sim\frac{1}{5}$ kpc) indicates that the gas is a collisional
fluid with an isotropic pressure tensor. The short sound crossing time
($<\sim 10^7$ yr) further indicates that, to a good approximation, the
gas is in hydrostatic equilibrium. Any streaming motions due to
cooling flows or supernova-driven winds should be very subsonic in
massive ellipticals; for a recent review see Sarazin (1997).

The traditional method for obtaining the masses of ellipticals from
X-rays assumes spherical symmetry and that the X-ray emission arises
from a single-phase, non-rotating, ideal gas in hydrostatic
equilibrium (Fabricant, Lecar, \& Gorenstein 1980),
\begin{eqnarray} 
M_{\rm grav}(<r) & \propto & rT(r)\left( \frac{d\ln\rho_{g}}{d\ln r} +
\frac{d\ln T}{d\ln r}\right), \label{eqn.sph}
\end{eqnarray} 
where $\rho_{g}$ is the mass density of the hot gas. The temperature
profile, $T(r)$, figures prominently in this equation, and has
historically hindered reliable mass determinations for most
ellipticals. Even at present accurate temperature profiles exist for
only a few galaxies and interpretations of these is hampered by the
possibilities of cooling flows and multi-temperature gas (Buote \&
Fabian 1997). Moreover, if $\vec{B}$ fields are important then an
additional term needs to added to equation (\ref{eqn.sph}).  Clearly
it is important to find a way to determine $M_{\rm grav}$ in a manner
that is insensitive to $T(r)$ and other poorly known properties of the
gas.

\section{A Geometric Test for Dark Matter} 
\label{gt}

The shapes of the X-ray isophotes probe $M_{\rm grav}$ without
requiring the equation of hydrostatic equilibrium to be solved
explicitly. For a galaxy of arbitrary shape the geometric properties
of the hydrostatic equation, $\nabla p_g = -\rho_g\nabla\Phi$, require
that the gas pressure, $p_g$, density, and gravitational potential,
$\Phi$, have the same shapes in three dimensions (3-D) independent of
the temperature profile of the gas; this can be shown by taking curls
of the hydrostatic equation.  (One replaces $\Phi$ with the
appropriate effective potential, $\Phi_{eff}$, if the gas is rotating
appreciably.)  If the gas is also adequately described by a
single-phase, $p_g = p_g(\rho_g,T)$, then $T$ has the same 3-D shapes
as $p_g$, $\rho_g$, and $\Phi$.  Since the X-ray emissivity is a
function only of these quantities, $j_x\propto\rho_g^2\Lambda(T)$
(where $\Lambda(T)$ is the intrinsic plasma emissivity), we arrive at
the key property (Buote \& Canizares 1994, \S 3.1; 1996a, \S 5.1),

\medskip
\noindent {\bf X-Ray Shape Theorem} {\it The X-ray emissivity and
gravitational potential have the same 3-D shapes independent of the
temperature profile of the gas.}

This relation between the shapes of $j_x$ and $\Phi$ is more robust to
issues like $\vec{B}$ fields and cooling flows than is equation
(\ref{eqn.sph}). Although in principle this relation may be affected
by arbitrary $\vec{B}$ fields, it is unaffected if $p_{mag}\propto
p_g$ as is often assumed (e.g., Loeb \& Mao 1994). Similarly,
simple models of cooling flows with mass dropout (White \& Sarazin
1987) just add another term to $j_x$ which is only a function of
$\rho_g$ and $T$, and thus do not disturb the relation. A possible
area of concern for this relation (as well as for equation
\ref{eqn.sph}) is if the gas is strongly multi-phase (see \S
\ref{caveats}).

The X-ray Shape Theorem allows a robust ``Geometric Test'' for dark
matter in ellipticals and galaxy clusters (Buote \& Canizares 1994,
1996a). That is, the hypothesis that gravitating mass follows the
optical light can be rigorously tested by comparing the shape of the
potential, $\Phi_L$, constructed from a constant $M/L$ model with the
shape of $j_x$ obtained from X-ray imaging data. The shape of $\Phi_L$
depends on both the shape {\it and} concentration of $\rho_L$, the
mass density of the constant $M/L$ model.


\refstepcounter{figure}
\addtocounter{figure}{-1}

\begin{figure}
\label{fig.pot}
\centerline{\hspace{0cm}\psfig{figure=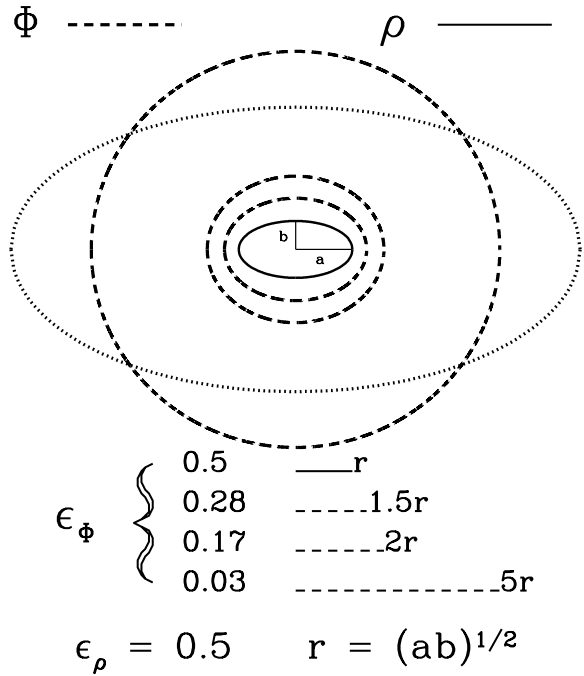,width=0.69\textwidth,angle=0}}
\caption{Potential shapes and mass concentration.}
\end{figure}

This dependence on mass concentration is illustrated in Figure
\ref{fig.pot} where the potential exterior to a thin ellipsoidal shell
with ellipticity, $\epsilon_{\rho}=0.5$, is shown. The exterior
isopotentials are confocal to the shell and thus their ellipticities
$(\epsilon_{\Phi})$ fall to nearly zero at only a few mean radii $(r)$
from the shell. (The dotted line is $\epsilon=0.5$ at $5r$ for
comparison.) This is an example of the general property that the
higher order potential multipoles rapidly decay with increasing
distance from a centrally concentrated mass and give way to the
spherical monopole. Hence, the ellipticity gradient of the X-ray
isophotes, which indicates a related gradient in the potential, probes
the radial mass distribution; see Buote \& Canizares (1996b) for an
application of this idea to clusters.

If the constant $M/L$ model cannot produce the X-ray isophote shapes
for a galaxy, then one can add a dark component until agreement is
achieved. Reasonable choices for this dark component are a halo having
either a NFW density profile (see Navarro, these proceedings) or a
Hernquist (1990) profile with the same ellipticity as the light. This
approach allows for a lower limit on the ratio of dark to luminous
matter, $M_{DM}/M_L$, to be determined independent of $T(r)$.

Finally, this ``Geometric Test'' can test the viability of alternative
gravity theories like MOND (Milgrom 1983), perhaps the most successful
of its kind.  Typically these alternative theories are devised to
explain radial manifestations of dark matter in galaxies (e.g., spiral
galaxy rotation curves).  Buote \& Canizares (1994) showed that
potential shapes in MOND are essentially identical to those in
Newtonian theory.  Hence, if a constant $M/L$ model in Newtonian
theory is unable to produce the observed X-ray isophote shapes of a
galaxy, neither will the corresponding model in MOND.

\section{Observational Results for the Geometric Test} 

\refstepcounter{figure}
\addtocounter{figure}{-1}

\begin{figure*}
\label{fig.n720}
\caption{NGC 720 PSPC (L) and HRI (R)}

\vskip 0.1cm

\parbox{0.49\textwidth}{
\centerline{\psfig{figure=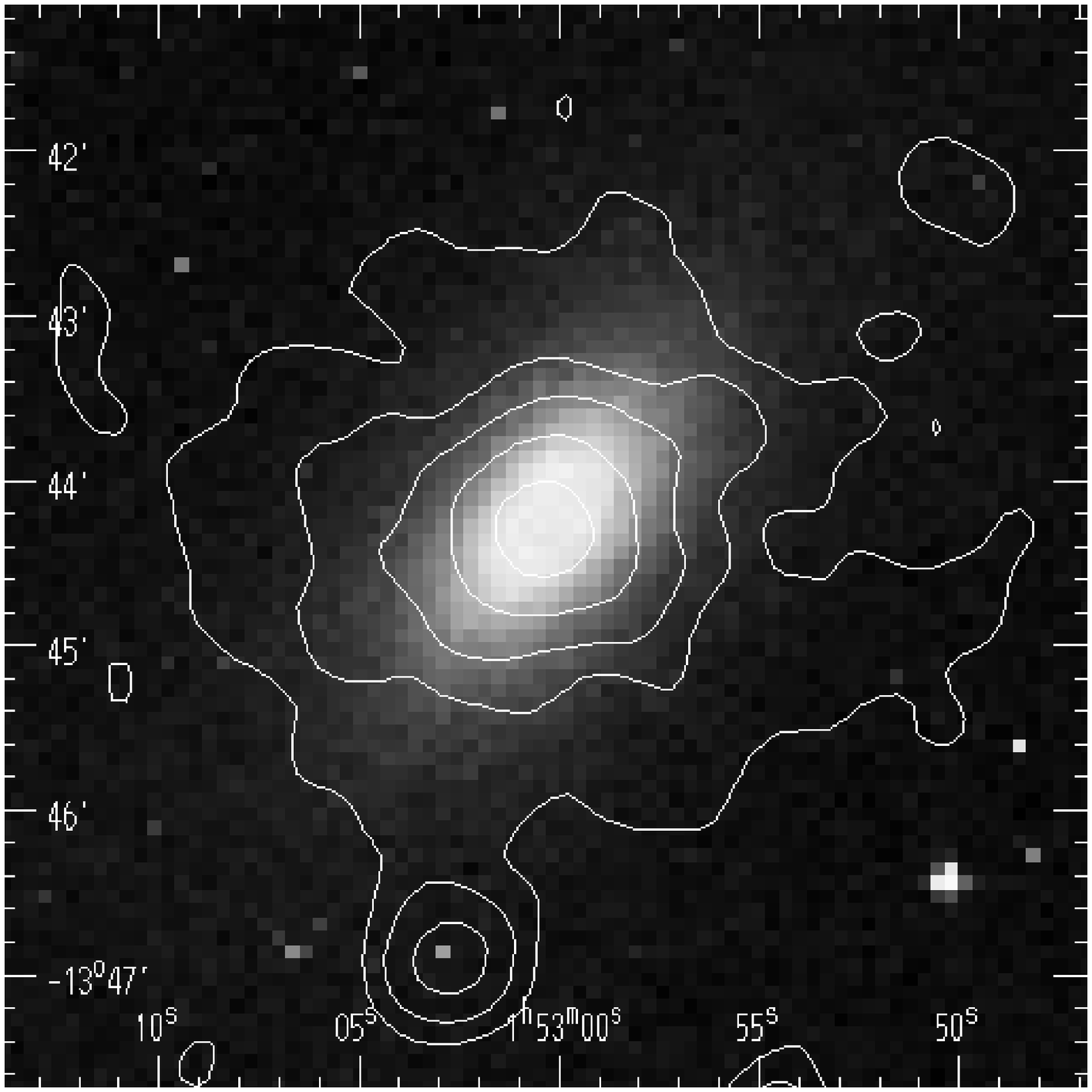,angle=0,height=0.3\textheight}}
}
\parbox{0.49\textwidth}{
\centerline{\psfig{figure=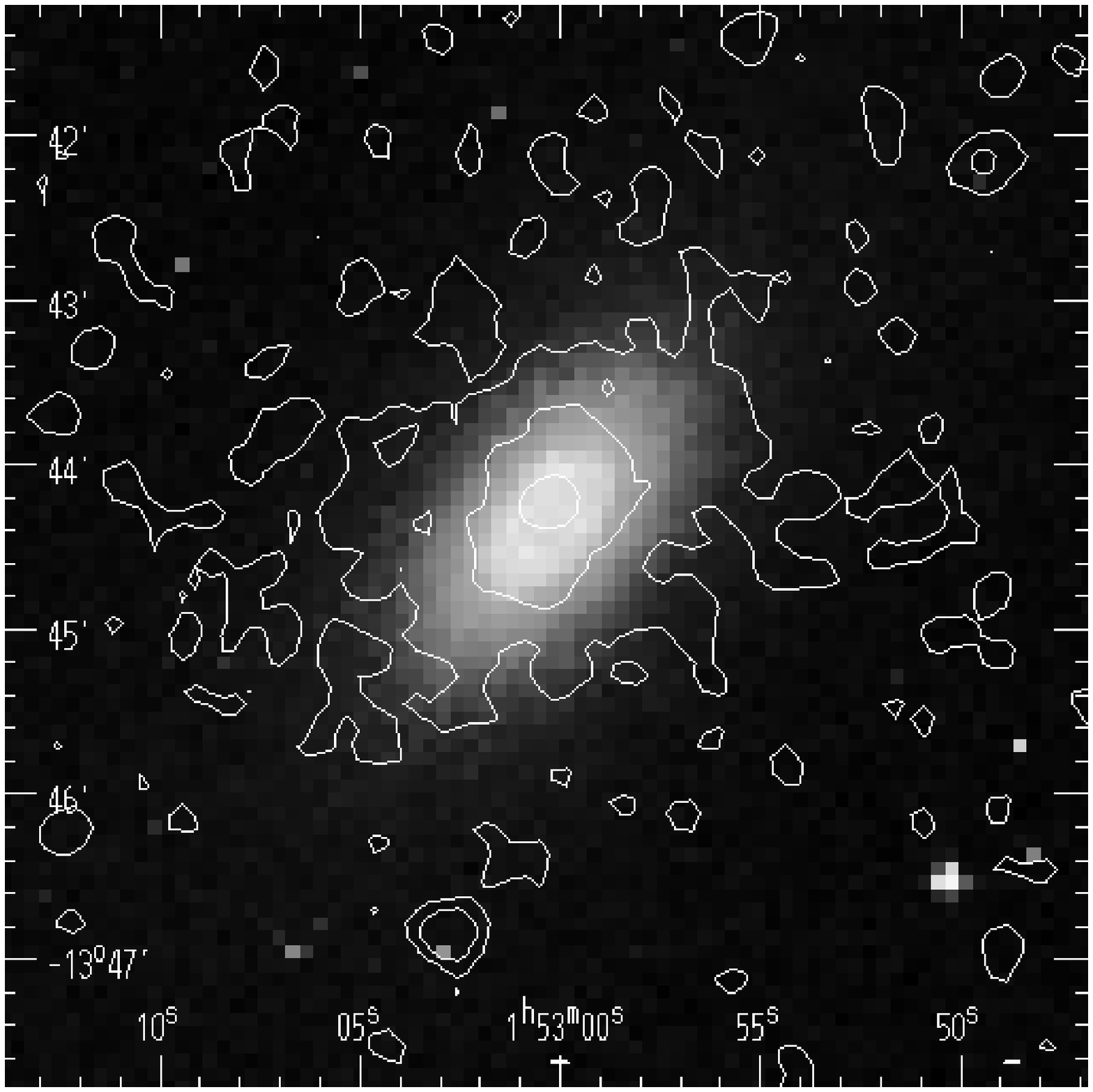,angle=0,height=0.3\textheight}}
}
\vskip 0.2cm
\centerline{\bf X-Ray and Optical Ellipticities} 

\parbox{0.49\textwidth}{
\centerline{\psfig{figure=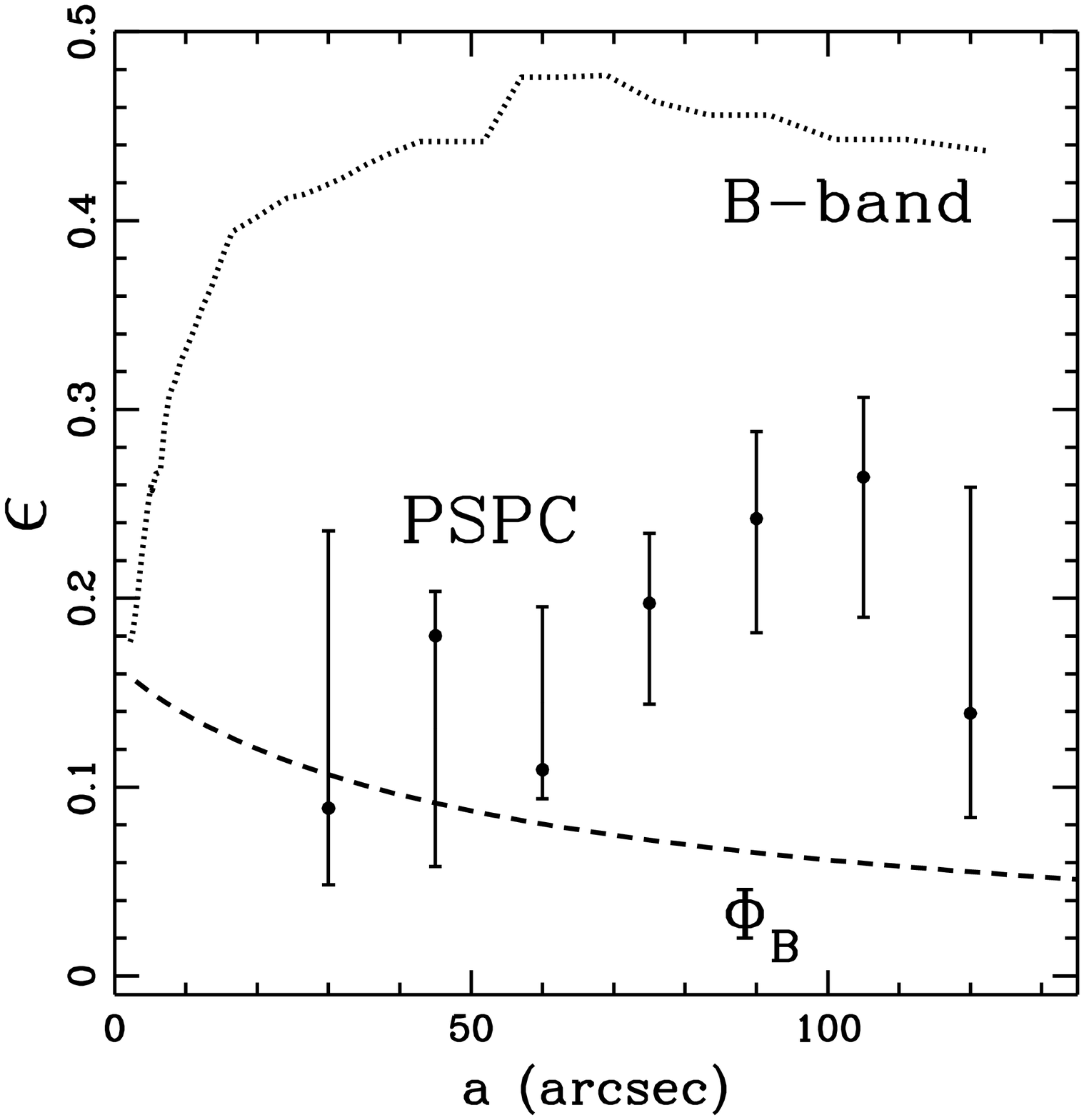,angle=0,height=0.3\textheight}}
}
\parbox{0.49\textwidth}{
\centerline{\psfig{figure=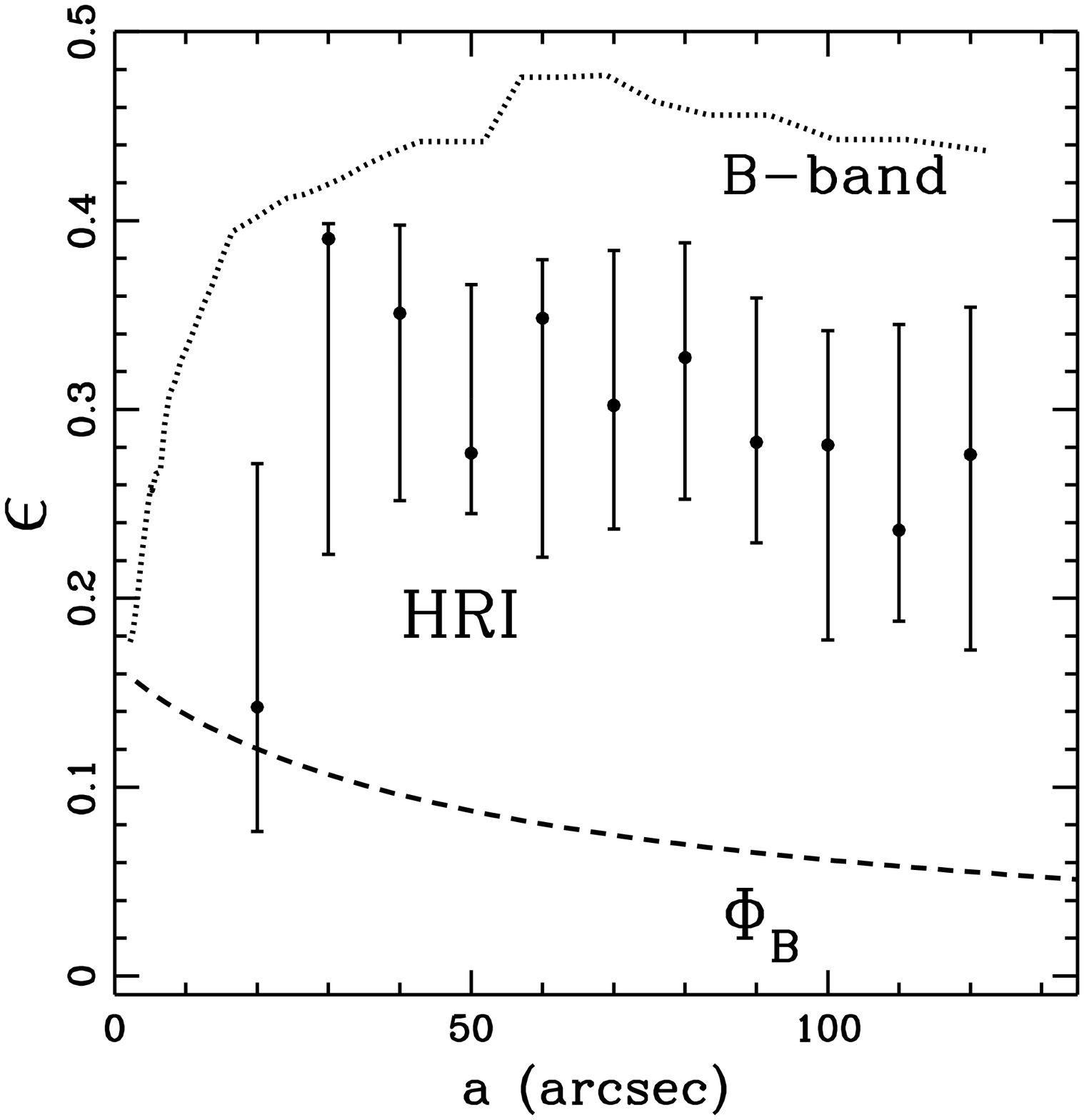,angle=0,height=0.3\textheight}}
}

\vskip 0.1 cm
\centerline{\bf Geometric Test for DM} 

\parbox{0.49\textwidth}{
\centerline{\psfig{figure=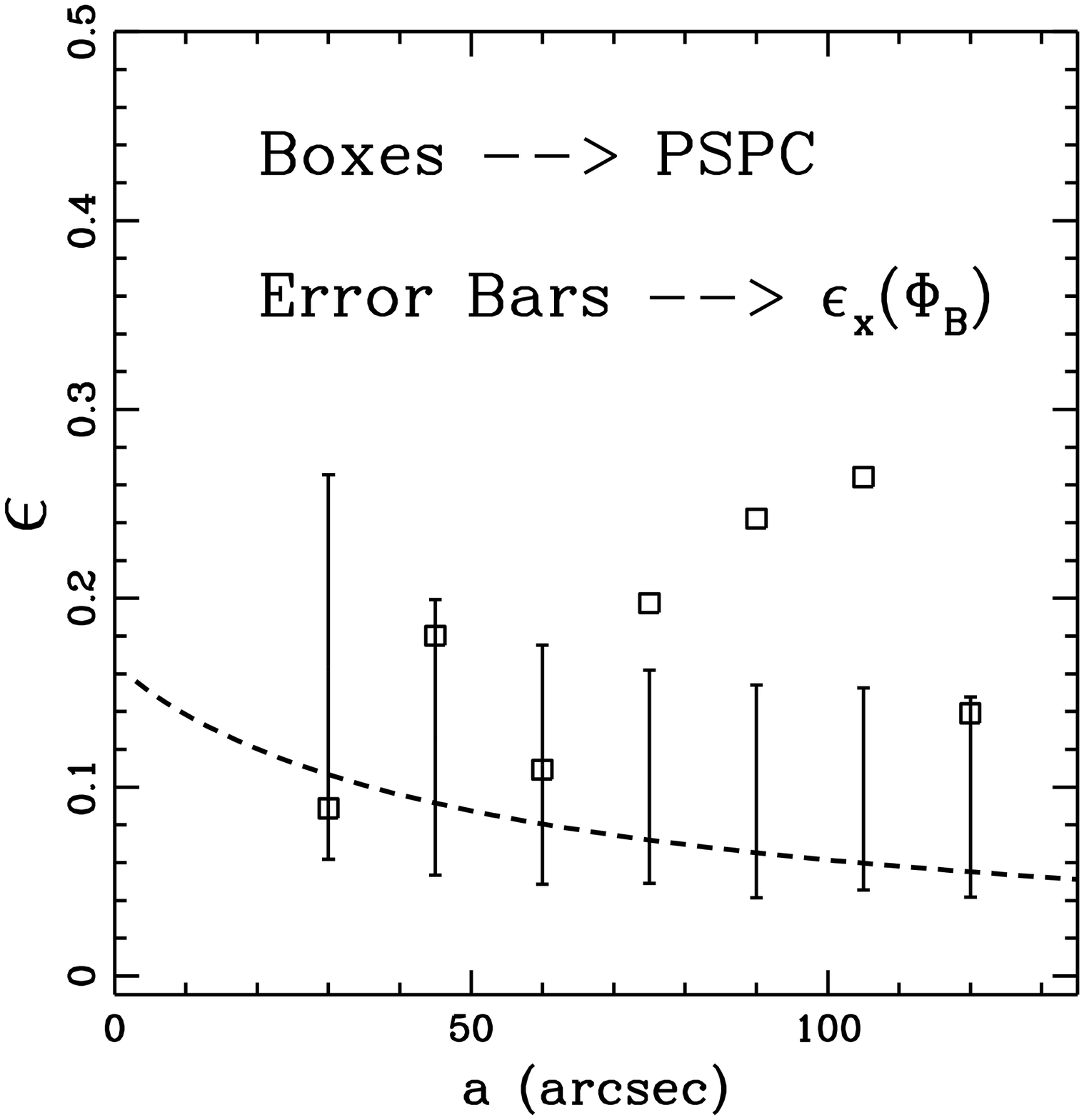,angle=0,height=0.3\textheight}}
}
\parbox{0.49\textwidth}{
\centerline{\psfig{figure=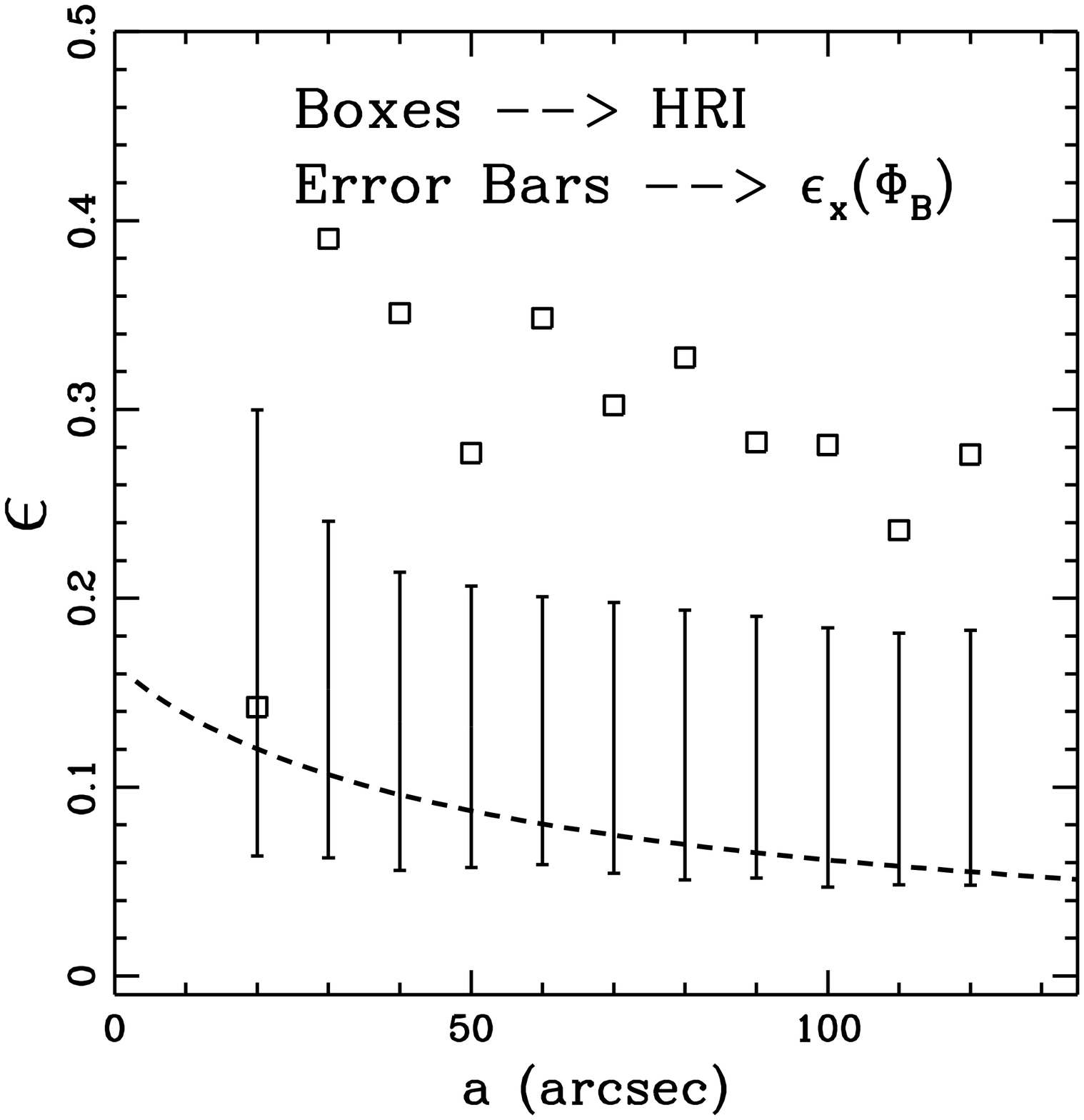,angle=0,height=0.3\textheight}}
}
\end{figure*}

The isolated E4 galaxy NGC 720 is the prototype for X-ray shape
analysis (see Buote \& Canizares 1994, 1996c, 1997; Romanowsky \&
Kochanek 1997). In Figure \ref{fig.n720} we show the X-ray contours of
the 20ks ROSAT PSPC and 57ks HRI observations of NGC 720 overlaid on
the digitized POSS image; $10^{\prime\prime}\sim 1h^{-1}_{75}$ kpc.
(The X-ray images have been smoothed for display with Gaussians having
half the widths of the respective PSFs; FWHM PSF is $\sim
30^{\prime\prime}$ for PSPC and is $\sim 4^{\prime\prime}$ for HRI.)
The X-ray isophotes are clearly flattened and the orientiations appear
to be offset from the optical major axis at large radii ($\sim
100^{\prime\prime}$). Here we concentrate on the ellipticity of the
X-ray surface brightness and defer discussion of the orientations to
\S \ref{pa}.

Because of the relatively low S/N data of the X-ray images, we use a
moment method to compute the ellipticity of the X-ray surface
brightness analogous to computing the principal moments of inertia
within an elliptical aperture (see Buote \& Canizares 1994).  The
X-ray ellipticity profiles are shown in Figure \ref{fig.n720} with
$1\sigma$ error bars from 1000 Monte carlo realizations of the images
(see Buote \& Canizares 1997); also displayed are the B-band isophotal
ellipticities. The ellipticity of the PSPC data is largest and most
significant for semi-major axes $a\sim
90^{\prime\prime}-100^{\prime\prime}$, while the HRI measures similar
ellipticity down to $a\sim 40^{\prime\prime}$.

By assuming $M\propto L_B$ we compute the potential $\Phi_B$; the
isopotential ellipticities appear in the figure.  The ellipticities of
$\Phi_B$ are considerably smaller than for the B-band light because
$L_B$ is highly centrally concentrated: $R_e\approx
50^{\prime\prime}$, core $\sim 4^{\prime\prime}$. That is, the
monopole dominates $\Phi_B$ for $r>\sim 10^{\prime\prime}$.  Since the
ellipticities of $\Phi_B$ fall well below those computed for the X-ray
data, the $M\propto L_B$ model would appear to fail. However, to
rigorously implement the Geometric Test we must deproject the X-ray
data to compare to the shape of $\Phi_B$.

The procedure we adopt, appropriate for the relatively low S/N X-ray
data ($\sim 1500$ counts), is to first jointly fit a simple model to
the radial surface brightness of the PSPC and HRI data (e.g., a
$\beta$ model). The best-fit model is deprojected to 3-D and assigned
the ellipticities of $\Phi_B$. By projecting this ellipsoidal model
back onto the sky plane (and adjusting the free parameters to maintain
a best fit of the radial surface brightness profile), we obtain a
$M\propto L_B$ model of the X-ray surface brightness. As above, we
perform 1000 Monte carlo realizations of this model and compute moment
ellipticities, $\epsilon_x(\Phi_B)$, analogous to the data to arrive
at the $1\sigma$ error bars in the bottom panels of Figure
\ref{fig.n720}.

The observed PSPC and HRI ellipticities exceed the
$\epsilon_x(\Phi_B)$ generated by the $M\propto L_B$ model at greater
than the $3\sigma$ level, thus indicating the need for flattened dark
matter.  Moreover, even if $L_B$ is allowed to have ellipticity $\sim
0.6$, consistent with the flattest ellipticals observed and expected
from stability arguments, this discrepancy cannot be entirely
resolved. Hence, in NGC 720 the dark matter must be both flattened and
more extended than $L_B$, a conclusion independent of the gas
temperature profile (Buote \& Canizares 1994, 1997).

\refstepcounter{figure}
\addtocounter{figure}{-1}

\begin{figure*}
\label{fig.other}
\caption{NGC 1332 (L) and NGC 3923 (R)}

\vskip 0.1cm

\parbox{0.49\textwidth}{
\centerline{\psfig{figure=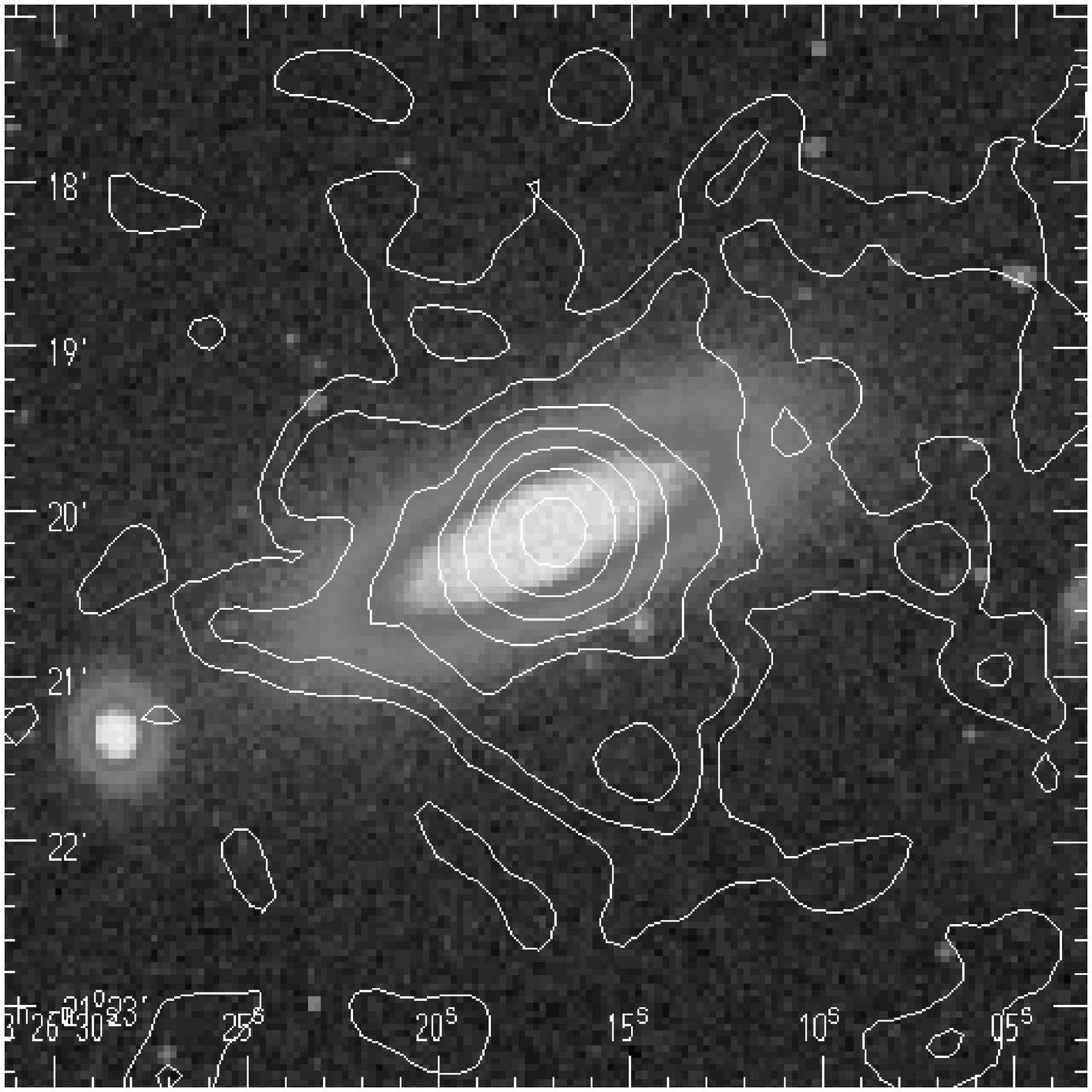,angle=0,height=0.3\textheight}}
}
\parbox{0.49\textwidth}{
\centerline{\psfig{figure=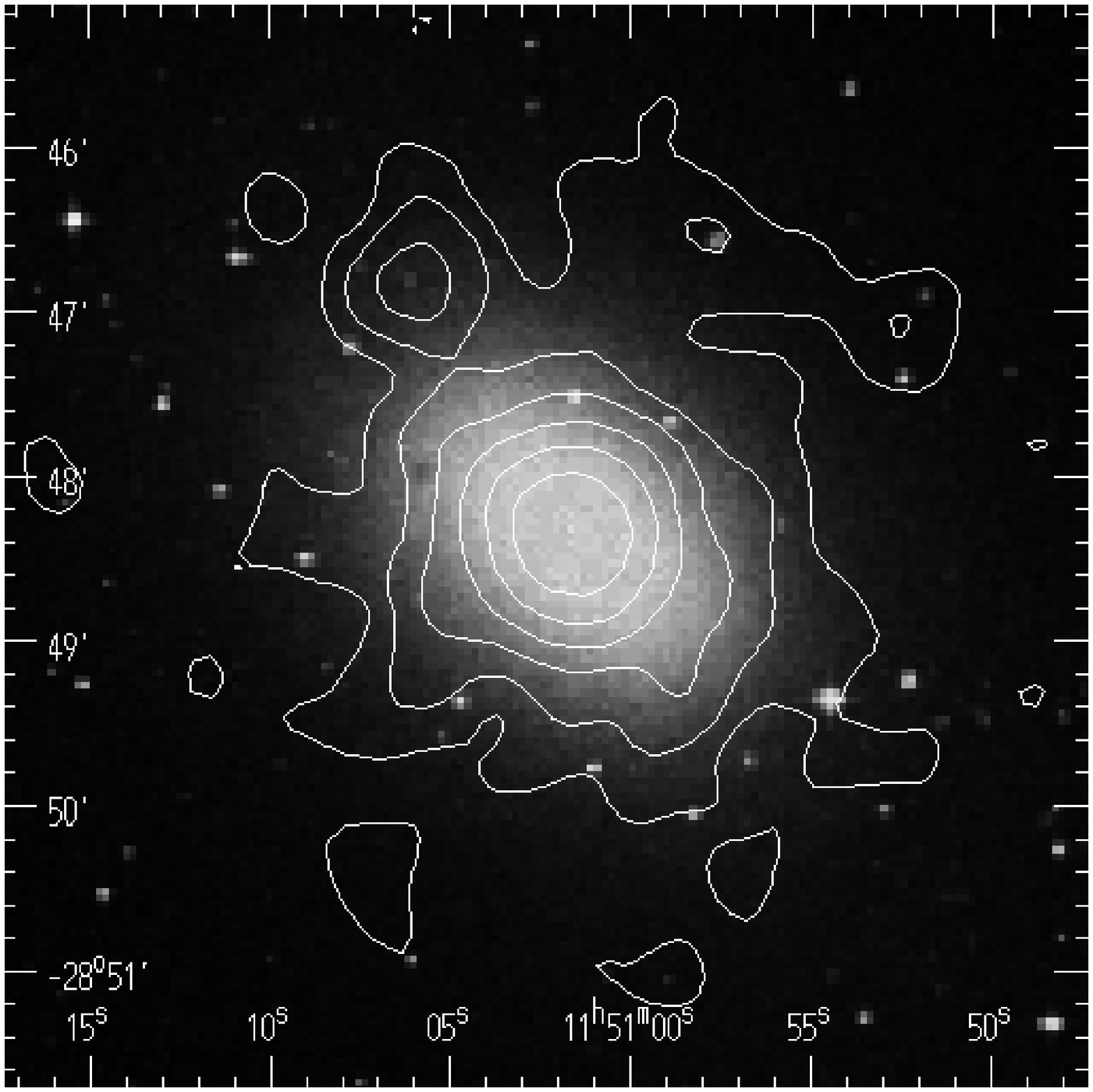,angle=0,height=0.3\textheight}}
}
\vskip 0.2cm
\centerline{\bf X-Ray and Optical Ellipticities} 

\parbox{0.49\textwidth}{
\centerline{\psfig{figure=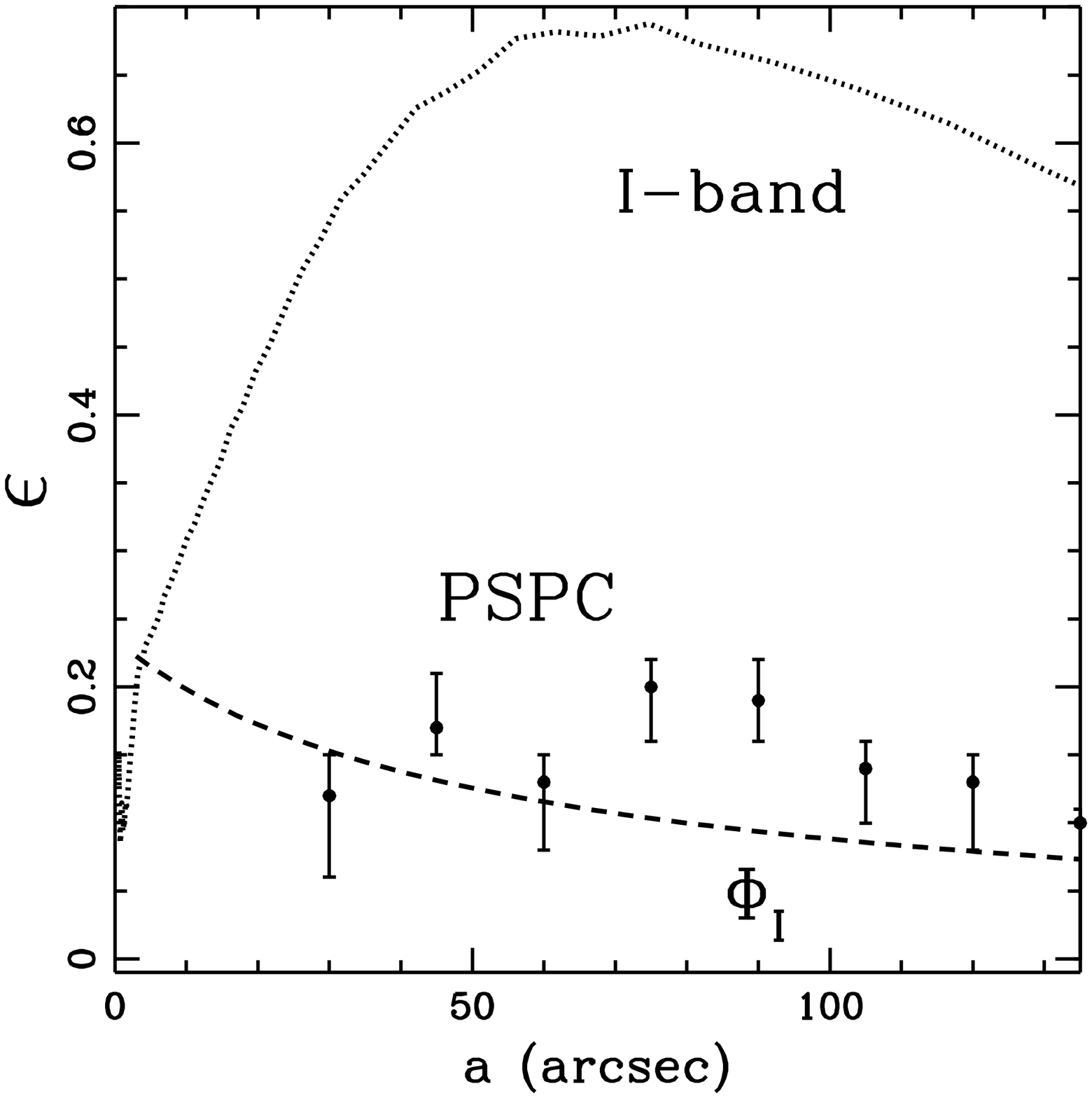,angle=0,height=0.3\textheight}}
}
\parbox{0.49\textwidth}{
\centerline{\psfig{figure=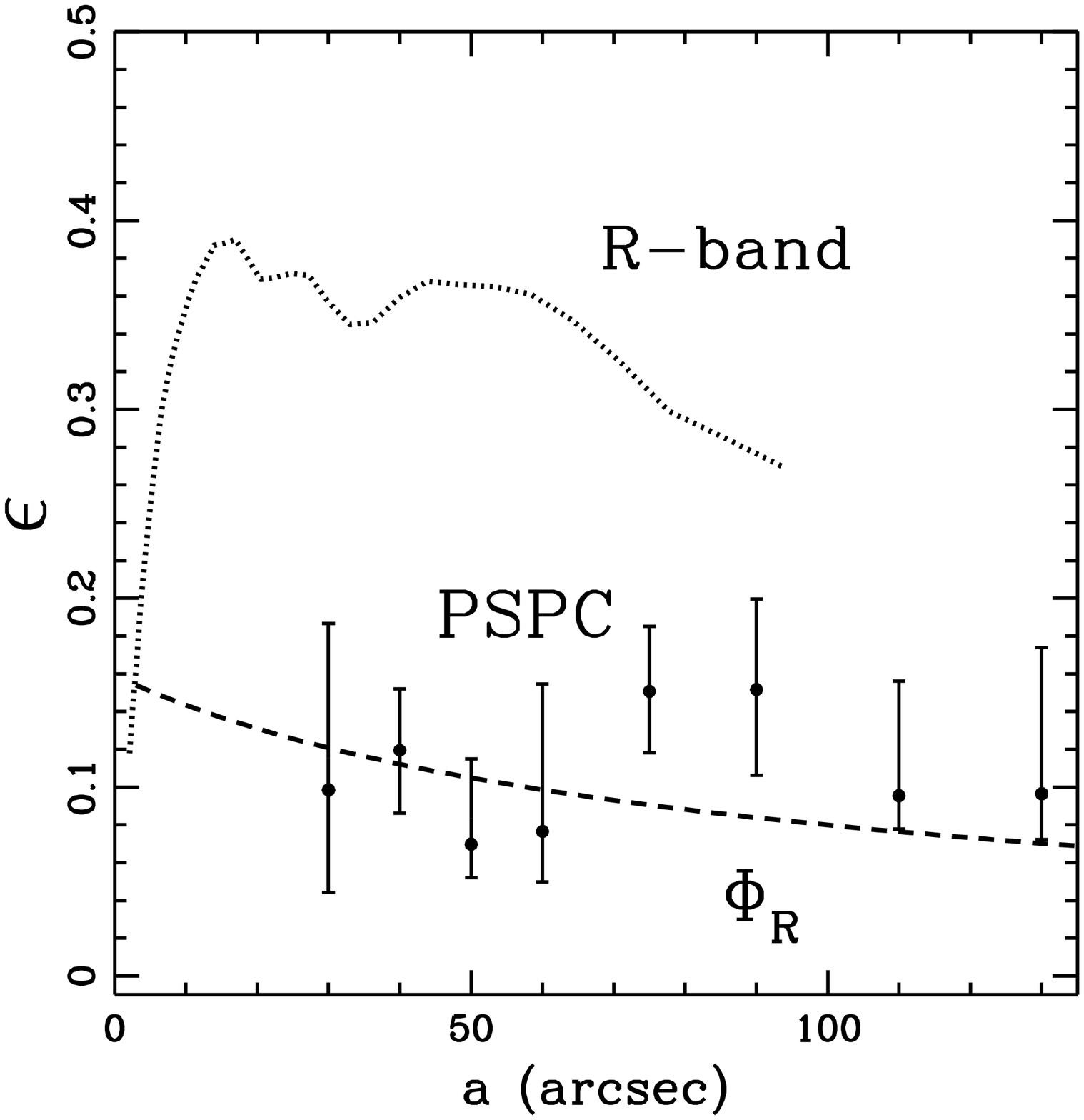,angle=0,height=0.3\textheight}}
}

\vskip 0.1 cm
\centerline{\bf Geometric Test for DM} 

\parbox{0.49\textwidth}{
\centerline{\psfig{figure=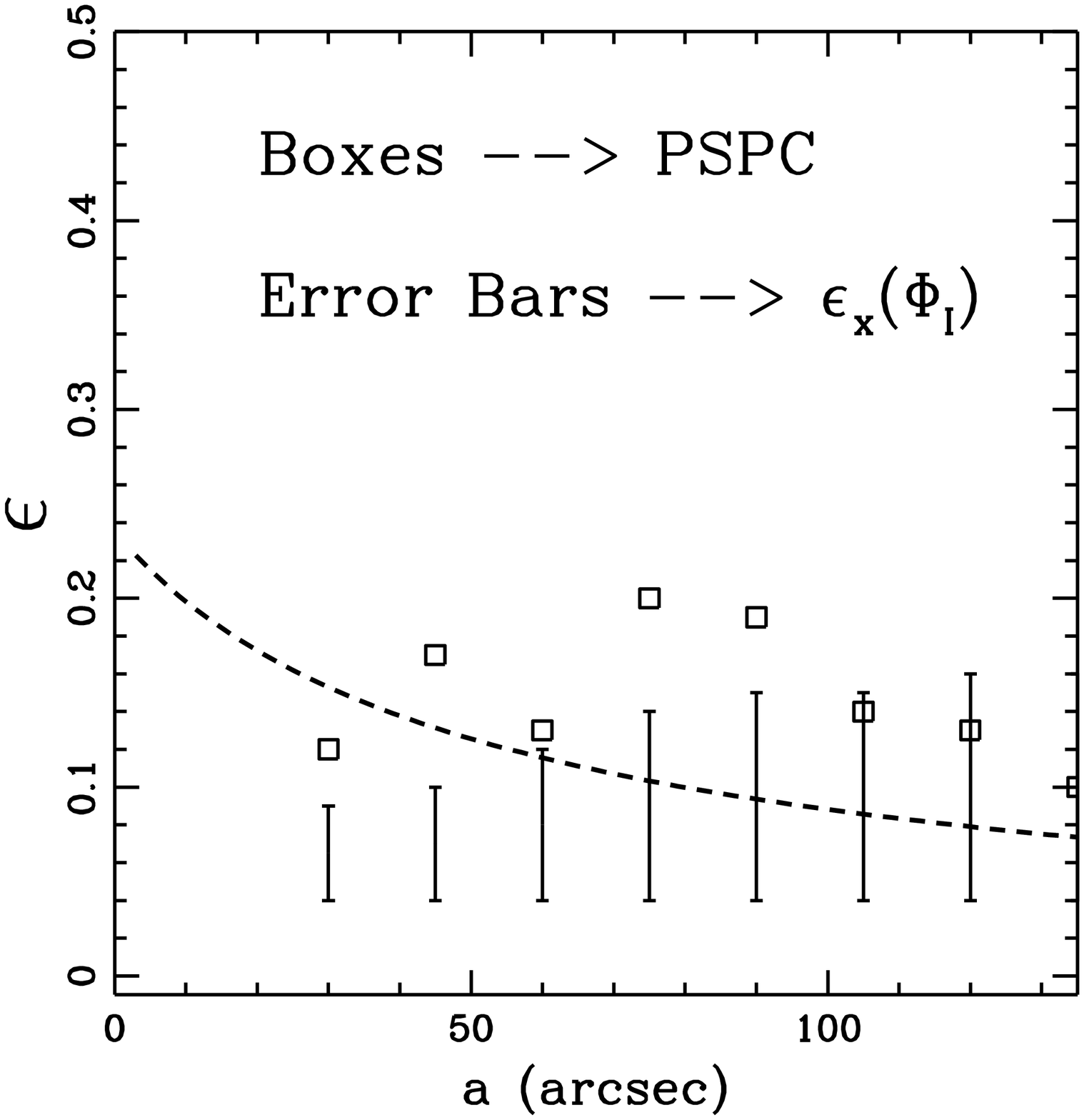,angle=0,height=0.3\textheight}}
}
\parbox{0.49\textwidth}{
\centerline{\psfig{figure=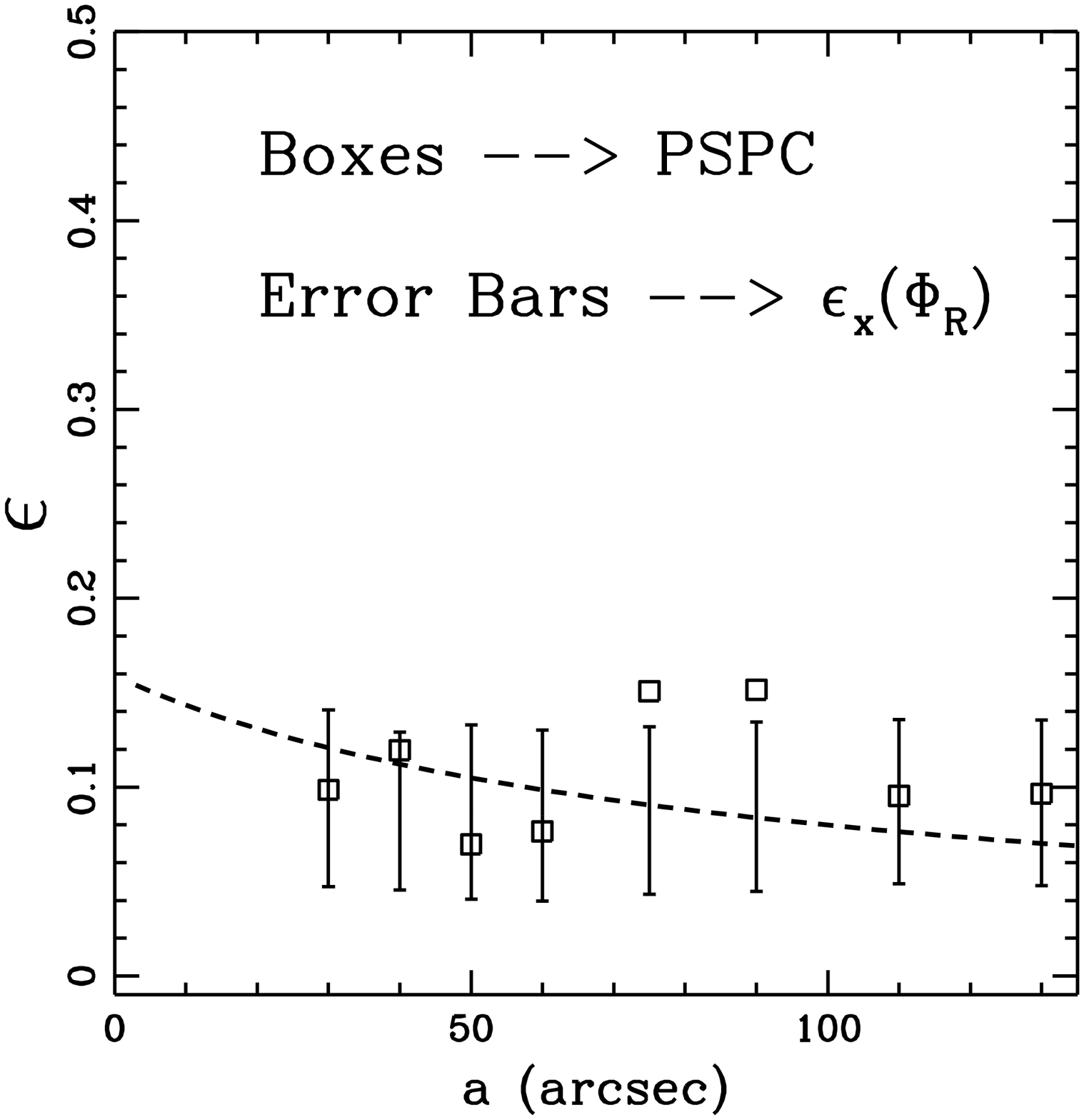,angle=0,height=0.3\textheight}}
}
\end{figure*}

We have also applied the Geometric Test to PSPC data of the fairly
isolated E7/S0 galaxy NGC 1332 (left panels of Figure \ref{fig.other})
and to PSPC data of the classic shell galaxy -- the isolated E4, NGC
3923 (right panels of Figure \ref{fig.other}).  Both of these galaxies
have lower S/N images than NGC 720 and have smaller measured X-ray
ellipticities. Similar to NGC 720, the quite flattened optical light
cannot produce flat enough X-ray isophotes. For NGC 1332, the observed
PSPC ellipticities exceed the $\epsilon_x(\Phi_I)$ generated by the
$M\propto L_I$ model at the $90\%$ level and, similar to NGC 720, both
flattened and extended dark matter are indicated (Buote \& Canizares
1996a). The $M\propto L_R$ model for NGC 3923 disagrees only at the
$80\%$ level. Although this latter case is marginal, the character of
the discrepancy is the same (i.e. flattened and extended dark matter
needed -- Buote \& Canizares 1997, to be submitted to ApJ).

\section{Detailed Hydrostatic Models} 

The Geometric Test robustly examines the need for dark matter and
provides some constraints on its shape and extent.  To obtain detailed
information on the range of allowed shapes and radial profiles for the
dark matter the hydrostatic equation must be solved explicitly.
Following the pioneering approach of Binney \& Strimple (1978), one
may solve the hydrostatic equation for the gas density by assuming a
single-phase, non-rotating ideal gas,
\begin{eqnarray}
\tilde{\rho}_{g} &  =  & \tilde{T}^{-1} \exp
 \left(-\Gamma\int\limits_0^{\vec{x}}\tilde{T}^{-1}\nabla\tilde{\Phi} \cdot
d{\vec x} \right), \label{eqn.he}
\end{eqnarray}
where $\Gamma = \mu m_p\Phi(0)/k_BT(0)$ and where, e.g.,
$\tilde{\rho}_{g} = \rho_g(\vec{x})/\rho_g(0)$. Starting with a model
for the mass distribution one computes $\Phi$, and then, in
conjunction with a measurement (or assumption) for the temperature
profile, one computes $\tilde{\rho}_{g}$. The model X-ray surface
brightness is then obtained by integrating $\tilde{\rho}_{g}^2$ along
the line-of-sight. This is appropriate since $\Lambda(T)$ convolved
with the ROSAT spectral response is effectively constant for the
relevant range of temperatures.

Although not immediately apparent from equation (\ref{eqn.he}),
constraints on the {\it shape} of the mass distribution using this
method are quite insensitive to temperature gradients (though not
independent of them like the Geometric Test). Strimple \& Binney
(1979) were the first to show that isothermal and adiabatic
($p_g\propto\rho_g^{5/3}$) temperature profiles lead to approximately
similar X-ray isophote shapes. Hence, an isothermal gas may be
reasonably assumed if one is only interested in measuring the shape of
the gravitating mass.

\begin{figure}
\label{fig.dmshape}
\centerline{\hspace{0cm}\psfig{figure=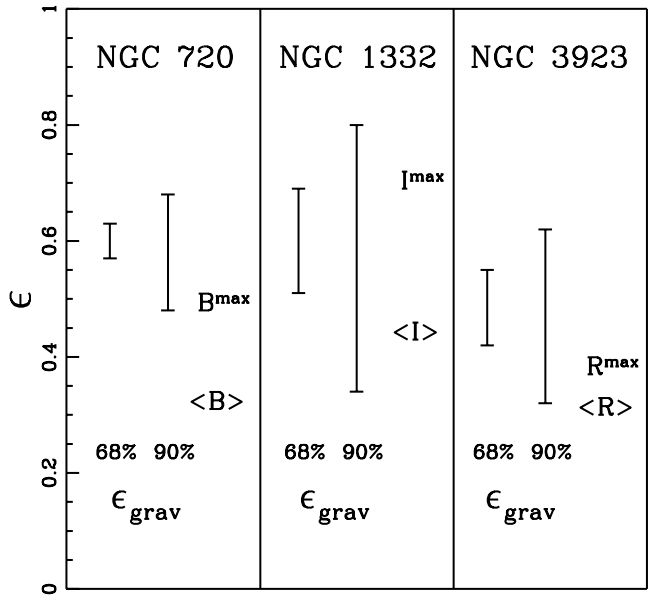,width=0.69\textwidth,angle=0}}
\caption{Ellipticity of Gravitating Mass}
\end{figure}

The results for the ellipticity of the gravitating matter in NGC 720,
NGC 1332, and NGC 3923 are shown in Figure \ref{fig.dmshape}. These
correspond to oblate spheroidal models with mass density, $\rho\sim
r^{-2}$, and for an isothermal gas (see Buote \& Canizares 1994,
1996a, 1997). (The results do not vary much for other models.)  The
intensity weighted optical ellipticity (e.g., $<B>$) and the maximum
value (e.g., $B^{max}$) are also shown. One can see that the
ellipticity of the gravitating matter is at least as large as that of
the light in all three cases, with some indication (especially NGC
720) that the dark matter is more elongated. 

Finally, the scale lengths of the gravitating matter are $\sim 5$
times that of the stellar luminosity for NGC 720, and $\sim 3$ times
for NGC 1332 and $\sim 2$ times for NGC 3923. These scale length
ratios are quoted for Hernquist models to insure a consistent
comparison between the light and mass, and the precise values depend
on the temperature profile.

\section{Caveats} 
\label{caveats}

The principal caveats associated with the analysis of X-ray isophote
shapes described above are, (1) multi-phase gas, (2) discrete sources,
(3) rotation, and (4) environmental effects.  There is new evidence
from ASCA that the hot gas in the brightest ellipticals consists of at
least two temperatures (Buote \& Fabian 1997).  The single-phase X-ray
shape analysis is insensitive to multiple gas phases if either (a) a
single phase dominates the emission, (b) (see \S \ref{gt}) the
emission is adequately described as a simple superposition of an
ambient phase and a mass dropout term (White \& Sarazin 1987), or (c)
the phases have similar spatial distributions so that the
emission-weighted (or mass-weighted) gas density and temperature are
good descriptions of the emissivity; i.e. $j_x \propto
\langle\rho_g\rangle^2\Lambda(\langle T\rangle)$. These mean
quantities appear to provide a good description of general multi-phase
cooling flows in clusters (Thomas, Fabian, \& Nulsen 1987).

Discrete sources should contribute to the soft X-ray emission of
early-type galaxies. In fact, the ASCA spectra of NGC 720, NGC 1332,
and NGC 3923 do require a second thermal component with $T>\sim 5$ keV
suggestive of discrete sources (Buote \& Canizares 1997; Buote \&
Fabian 1997).  However, the large inferred temperatures of these hard
components in these relatively low S/N galaxies may be a fitting
artifact since the highest S/N ellipticals have ``hard'' components
with $T<\sim 2$ keV which instead indicate another phase of hot gas
(Buote \& Fabian 1997).  Moreover, the spatial distribution of the
discrete component has not been accurately measured. However,
assuming the discrete sources contribute to 20\% of the ROSAT
emission of NGC 720 and that they are distributed like the optical
light, Buote \& Canizares (1997) have shown that the inferred
ellipticity of the gravitating mass decreases by only $\sim 0.05$.

It is unlikely that rotation plays an important role in the dynamics
of NGC 720 and NGC 3923 since the stellar rotation is
negligible. Theoretically, even without strong stellar rotation one
may (Kley \& Mathews 1994) or may not (Nulsen, Stewart, \& Fabian
1984) expect a rotating cooling flow to develop depending on whether
angular momentum of the gas is conserved. Highly flattened X-ray
isophotes indicative of a cooling disk have not been
observed. Finally, it is unlikely that environmental effects
substantially distort the X-ray isophotes in the isolated ellipticals
studied since for $r<\sim 100^{\prime\prime}$ no large asymmetries
indicative of ram pressure or tidal distortions are observed.

\section{X-ray position angle twist in NGC 720.} 
\label{pa}

\refstepcounter{figure}
\addtocounter{figure}{-1}

\begin{figure*}
\label{fig.pa}
\parbox{0.49\textwidth}{
\centerline{\psfig{figure=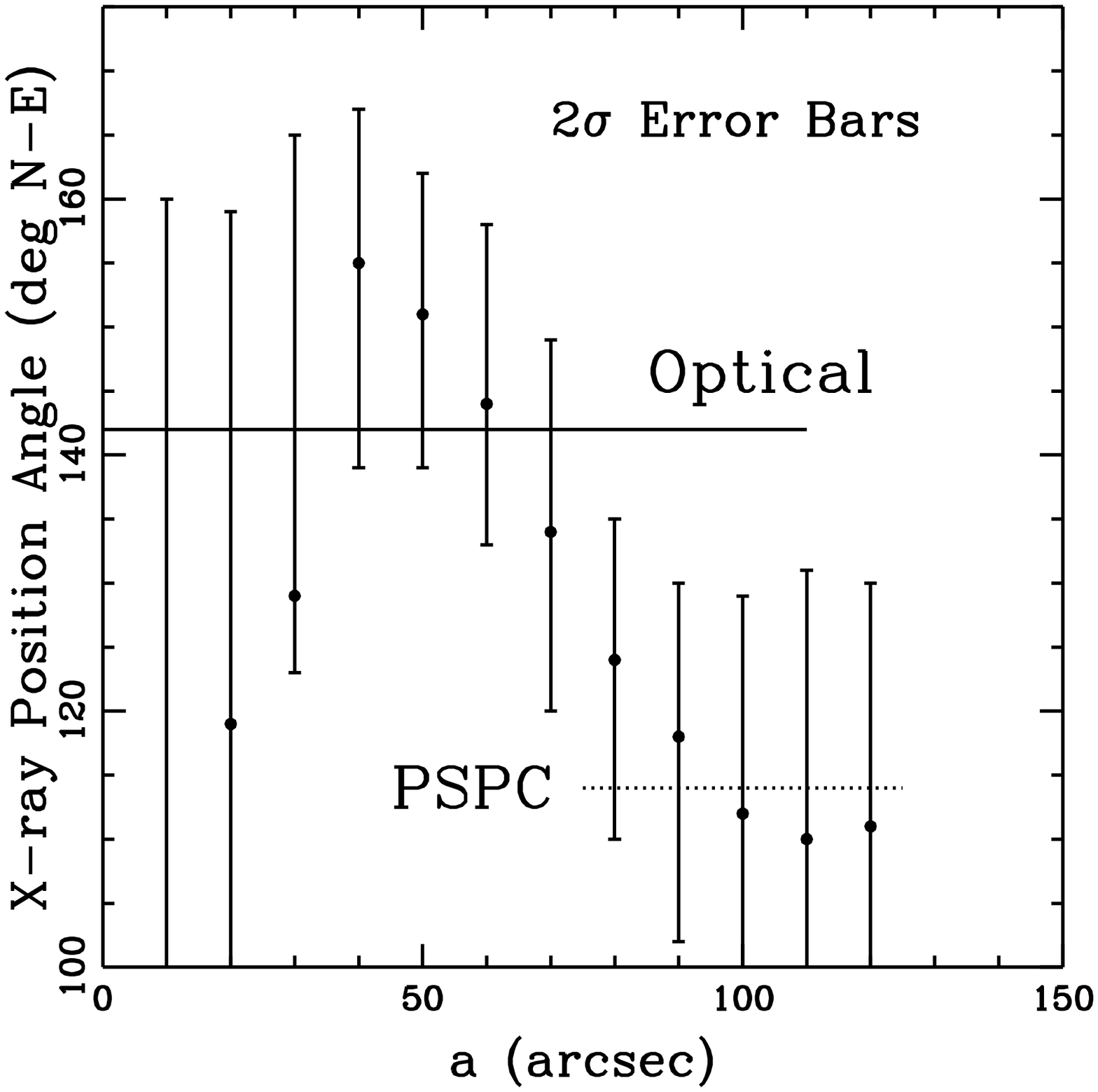,angle=0,height=0.3\textheight}}
}
\parbox{0.49\textwidth}{
\centerline{\psfig{figure=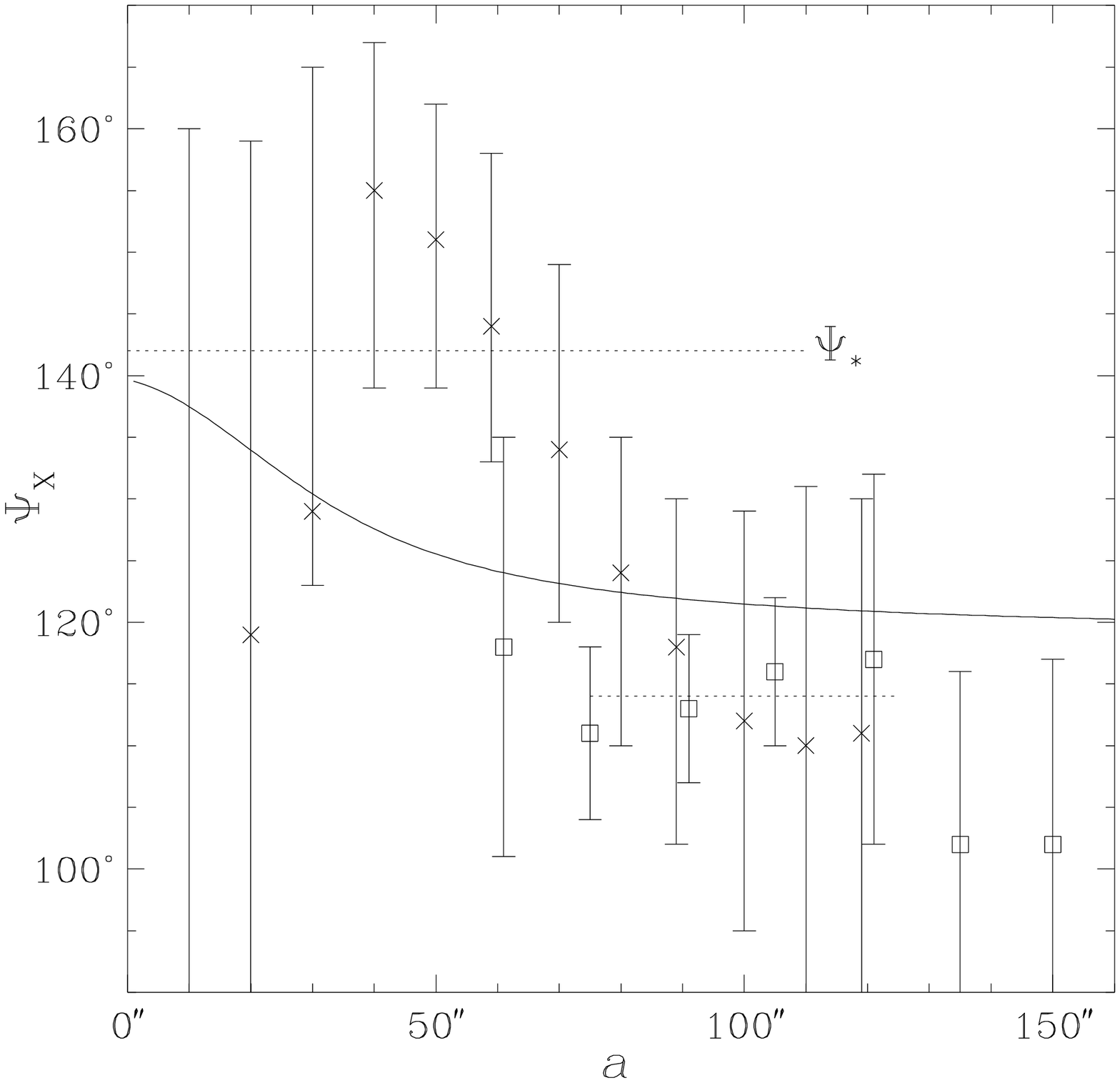,angle=0,height=0.3\textheight}}
}
\caption{X-Ray Position Angle Twist in NGC 720}
\end{figure*}

The major axes of the X-ray isophotes of NGC 720 (Figure
\ref{fig.n720}) are offset from the optical major axes. In the left
panel of Figure \ref{fig.pa} we show the position angle (PA) profile
of the HRI data with the PAs of the PSPC and optical data indicated
for comparison (Buote \& Canizares 1996c). For $r<\sim
60^{\prime\prime}$ the X-ray isophotes are aligned with the optical
and then twist away to a maximum misalignment of $\sim 30$ degrees at
$r\sim 100^{\prime\prime}$. This misalignment provides further
geometric evidence for dark matter. Buote \& Canizares (1996c)
attempted to explain this PA twist with a simple triaxial halo model
where the axial ratio varied as a power law in radius but were unable
to simultaneously produce the large twist while maintaining the
relatively large X-ray ellipticities.

Romanowsky \& Kochanek (1997) have recently examined more detailed
physical triaxial models to try to explain the NGC 720 X-ray
data. Although the best model of Romanowsky \& Kochanek (solid line in
right panel of Figure \ref{fig.pa}) is able to produce a large twist,
the twist occurs at very small radii and is thus unable to explain the
abrupt twist at $\sim 60^{\prime\prime}$ seen in the HRI data. An
intrinsic misalignment of the stars and dark matter halo may be
implicated.

\section{Conclusions} 

X-ray isophote shapes probe both the shape and concentration of
gravitating mass essentially independent of the temperature profile of
the gas. (The Geometric Test is completely independent of $T(r)$.) For
the early-type galaxies studied (NGC 720, NGC 1332, NGC 3923), dark
matter is required to explain the elongated isophotes. (MOND is unable
to escape this manifestation of dark matter.) The dark matter is at
least as flattened as the optical light and is also more extended for
each of the galaxies. The X-ray position-angle twist in NGC 720
appears to suggest an intrinsic misalignment of stars and dark matter
halo rather than a strongly triaxial system.

This X-ray shape analysis is applicable to isolated early-type
galaxies with typical $\log_{10}L_x/L_B>\sim$ $-2.7$ (i.e. X-rays from
hot gas). The next generation X-ray satellites soon to be flown, AXAF
and XMM, will have the spatial resolution and effective area to
precisely map X-ray isophote shapes for a statistically large sample
of galaxies. X-rays are the most promising means on the horizon for
obtaining a large sample of gravitating mass shapes and radial mass
profiles of early-type galaxies which should significantly enhance our
understanding of the structure and formation of these systems (e.g.,
Sackett 1996; de Zeeuw 1997).

%
%

%

\end{document}